\documentclass[runningheads]{llncs}
\usepackage{graphicx}
\usepackage{color}
\usepackage{colortbl}
\usepackage{comment}

\begin{document}
\title{What are the Actual Flaws in Important Smart Contracts (and
  How Can We Find Them)?}

\author{Alex Groce\inst{1} \and Josselin Feist\inst{2} \and Gustavo
 Grieco\inst{2} \and Michael Colburn\inst{2}}

\authorrunning{A. Groce, J. Feist, G. Grieco, and M. Colburn}
\titlerunning{What are the Actual Flaws in Important Smart Contracts?}

\institute{Northern Arizona University, Flagstaff AZ 86011 \and Trail of Bits, New York NY 10003}

\maketitle

\newcommand{\alex}[1]{\textbf{\color{blue} alex: #1}}
\newcommand{\josselin}[1]{\textbf{\color{red} jo: #1}}
\newcommand{\gustavo}[1]{\textbf{\color{cyan} gus: #1}}

\definecolor{Gray}{gray}{0.9}

\begin{abstract}
An important problem in smart contract security is understanding the likelihood and criticality of discovered, or potential, weaknesses in contracts. In this paper we provide a summary of Ethereum smart contract audits performed for 23 professional stakeholders, avoiding the common problem of reporting issues mostly prevalent in low-quality contracts. These audits were performed at a leading company in blockchain security, using both open-source and proprietary tools, as well as human code analysis performed by professional security engineers. We categorize 246 individual defects, making it possible to compare the severity and frequency of different vulnerability types, compare smart contract and non-smart contract flaws, and to estimate the efficacy of automated vulnerability detection approaches. 
\end{abstract}

\section{Introduction}

Smart contracts are versatile instruments that can not only facilitate and verify transactions in financial services, but also track the movement of physical goods and intellectual property. Security and correctness are essential for smart contract technology, because contracts possess the authority to allocate high-value resources between complex systems and are, for the most part, autonomous. 

Security researchers have worked to describe vulnerabilities and produce tools that find flaws in smart contracts, but most of the discussions of such flaws concentrate on a small number of actual exploits \cite{spank,DAO}. Moreover, many studies examine \emph{all} the contracts on a blockchain or focus on ``popular'' \cite{measurepop} contracts, but these contracts often are produced by development efforts where security and correctness are not prioritized. While informative, these analyses do not represent the contracts that are likely to become the infrastructure of a smart-contract future.

A better alternative for understanding smart contract flaws is to analyze bugs discovered during professional security audits. Early investors in smart contracts expose themselves to risks that could be devastating if the code is insecure or incorrect. Given these consequences, it is more likely that an initial effort is made to produce correct code. Therefore, flaws discovered during paid security audits provide a better ground truth for recommending ways to improve smart contract security. This paper presents an analysis of the types of flaws detected in 23 Solidity/Ethereum \cite{buterin2013whitepaper,wood2014yellow} smart contract audits performed by Trail of Bits (\url{https://trailofbits.com}), a leading company in the field.

\section{Related Work}

To our knowledge, \emph{no} previous work reports flaws detected in paid security
audits of important smart contracts.  We have not even found any manual
examination of large numbers of smart contracts with reasonable
criteria for removing uninteresting contracts (which would ensure quality
analysis).
However, there are other
important efforts to classify or describe smart contract flaws.
Atzei, Bartoletti, and Cimoli produced a taxonomy of possible
\emph{attacks} on smart contracts, with examples of actual exploit
code \cite{SurveyAttacks}.  Their categories have some overlap with
those used in this paper, but are more focused on specific-exploit
patterns and exclude some types of flaws that are not tied to a
specific attack.  We believe that every category present in
their taxonomy is also represented by at least one finding in our
set.  Their purpose is largely orthogonal to ours and presents a
useful alternative view of the topic, but one based more on
speculation about exploits than on concrete data about the prevalence and
seriousness of flaws in real contracts.   Mense and Flatscher
\cite{Mense} combine a summary of known vulnerability types with a
simple comparison of then-available tools, while Saad et
al. \cite{ExploreAttackSurface}  expand the scope of analysis to
general blockchain attack surfaces, but provide a similar
categorization of smart contract vulnerabilities.   Dika's thesis also
\cite{dika2017ethereum} provides another, earlier, summary of
vulnerability types, analyses, and tools. In general, the types of
flaws discussed in these works are a subset of those we discuss below.

Perez and Livshits
provide a (provocatively titled) analysis of actual executed exploits
on 21K contracts reported in various academic papers, which
provides a useful additional perspective, but they use a very different
data set with purposes almost completely unrelated
to ours \cite{DoesAnyoneCare}.   They find that,
while reentrancy is the most dangerous category of problem (over 65\%
of actual exploits in the wild), even reentrancy exploits have
resulted in loss of less than 10K Ether to date.  The relatively small
size of exploits to date vs. potential future losses affirms that
information about undetected flaws in audited
high-value, high-visibility contracts is important to the community.  

Smart contract analysis/verification research often touches on the
topic of expected vulnerabilities
\cite{slitherpaper,securify,smartcheck,Brent2018VandalAS,ethertrust,mythril-code,teether,oyente,smartanvil,grishchenko2018semantic,ethertrust,maian},
but this research is, to our knowledge, always based on author perceptions of
threats, not statistical inference from close examinations of
high-quality/critical contracts.

\section{Summary of Findings}

The results below are based on 23 audits performed by Trail of Bits.
Of these, all but five are public, and the reports are available
online \cite{reports}.  The number of findings per audit ranged from 2-22, with a median and mean of 10 findings.
Reports ranged in size from just under
2K words to nearly 13K words, with a total size of over 180K
words. 
It is also worth mentioning that each audit focused on a code-base that has between one to a few dozen of contracts that Trail of Bits reviewed manually and using automated tools.  The total number of audited contracts is thus considerably more than 23 (some individual audits covered more than 23 contracts).


The time allotted for audits ranged from one person-week to twelve
person-weeks, with a mean of six person-weeks and a median of four
person-weeks.  The audits were prepared by a total of 24 different auditors, with most audits prepared by multiple individuals (up to
five).  The mean number of authors was 2.6, and the median was three.  The most audits in which a single author participated was
12, the mean was 3.2; the median was only two audits.  In general,
while these audits are all the product of a single company, there is
considerable diversity in the set of experts involved.

Most of these assessments used static and dynamic analysis tools in addition to
manual analysis of code, but the primary source of findings was manual.  In
particular, a version of the Slither static analyzer \cite{slitherpaper} which included
a number of detectors not available in the public version, was applied to many of the contracts.  In some cases,
property-based testing with Echidna \cite{echidna-code} and symbolic
analysis with Manticore \cite{manticorepaper,manticore-code} were also applied to
detect some problems.  Only two audits did not use automated tools.  Sixteen of the audits made use of Slither,
sixteen made use of Manticore, and thirteen made use of Echidna.  However,
when Slither was used in audits, it was usually used much
more extensively than Manticore or Echidna, which were
typically restricted to a few chosen properties of high interest.
Only four findings are explicitly noted in the findings as produced by a
tool, all by Slither.  However, other findings may have resulted from
automated analyses in a less explicit fashion.

\subsection{Smart Contract Findings}

Our analysis is based on 246 total findings. Tables
\ref{tab:overallcatpercent} and \ref{tab:findway} summarize information on these
findings (Table \ref{tab:findway}). 
Each flaw is classified according to its severity,
considering the potential impact of the exploit to be:
\begin{itemize}
    \item \emph{High} if it affects a large numbers of users, or has serious legal and financial implications;
    \item \emph{Medium} if it affects individual users' information, or has possible legal implications for clients and moderate financial impact;
    \item \emph{Low} if the risk is relatively small or is not a risk the customer has indicated is important;
    \item \emph{Informational} if the issue does not pose an immediate risk, but is relevant to security best practices.
\end{itemize}

Another important property of each finding is how difficult it is to exploit:
\begin{itemize}
    \item \emph{Low} for commonly exploited flaws where public tools exist or exploitation can be easily automated;
    \item \emph{Medium} for flaws that require in-depth knowledge of a complex system;
    \item \emph{High} for flaws where an attacker must have privileged insider access to the system, or must discover other weaknesses, for exploitation.
\end{itemize}


The findings categories are sorted by the
frequency of severity counts; ties in the high-severity findings count are
broken by counting medium-severity findings, and further ties are broken by
low-severity findings.  Appendix A shows exact
counts for categories and severities/difficulties. Raw data is
also available~\cite{contractrepo}.

The categories in these tables are
generally the categories used in the audit reports submitted
to clients, but in some cases we have corrected obviously incorrect
categories given the continuous evolution of the security landscape for smart contracts.
Additionally,  we have introduced a few new categories in cases where
findings were clearly placed in a category of dubious relevance due
to the lack of a suitable category.  The most significant systematic change is
that we separated \emph{race conditions} and \emph{front-running} from
all other timing issues, due to 1) the large number of race conditions
relative to other timing issues; 2) the general qualitative
difference between race conditions and other timing-based exploits
(e.g., there is a large literature addressing detection and
mitigation of race conditions specifically); and 3) the specific relevance of
front-running to smart contracts.  Our analysis calls special
attention to findings classified as {\bf high-low}, that is {\bf high
  severity} and {\bf low difficulty}.  These offer attackers
an easy way to inflict potentially severe harm.  There were
27 high-low findings, all classified as one of eight categories:  data validation, access controls,
numerics, undefined behavior, patching, denial of service,
authentication, or timing.

\begin{table}[t]
  \begin{tabular}{lr|r|rrrrr|rrrr}
& & & \multicolumn{5}{c}{Severity} & \multicolumn{4}{|c}{Difficulty} \\
Category & \% & High-Low & High & Med. & Low & Info. & Und. & High & Med. & Low & Und.\\
\hline
\rowcolor{Gray}
data validation & 36\% & 11\% & 21\% & 36\% & 24\% & 13\% & 6\% & 27\% & 16\% & 55\% & 2\% \\
access controls & 10\% & 25\% & 42\% & 25\% & 12\% & 21\% & 0\% & 33\% & 12\% & 54\% & 0\% \\
\rowcolor{Gray}
race condition & 7\% & 0\% & 41\% & 41\% & 6\% & 12\% & 0\% & 100\% & 0\% & 0\% & 0\% \\
numerics & 5\% & 23\% & 31\% & 23\% & 38\% & 8\% & 0\% & 31\% & 8\% & 62\% & 0\% \\
\rowcolor{Gray}
undefined behavior & 5\% & 23\% & 31\% & 15\% & 31\% & 8\% & 15\% & 15\% & 8\% & 77\% & 0\% \\
patching & 7\% & 11\% & 17\% & 11\% & 39\% & 28\% & 6\% & 6\% & 11\% & 61\% & 22\% \\
\rowcolor{Gray}
denial of service & 4\% & 10\% & 20\% & 30\% & 30\% & 20\% & 0\% & 50\% & 0\% & 40\% & 10\% \\
authentication & 2\% & 25\% & 50\% & 25\% & 25\% & 0\% & 0\% & 50\% & 0\% & 50\% & 0\% \\
\rowcolor{Gray}
reentrancy & 2\% & 0\% & 50\% & 25\% & 25\% & 0\% & 0\% & 50\% & 25\% & 0\% & 25\% \\
error reporting & 3\% & 0\% & 29\% & 14\% & 0\% & 57\% & 0\% & 43\% & 29\% & 29\% & 0\% \\
\rowcolor{Gray}
configuration & 2\% & 0\% & 40\% & 0\% & 20\% & 20\% & 20\% & 60\% & 20\% & 20\% & 0\% \\
logic & 1\% & 0\% & 33\% & 33\% & 33\% & 0\% & 0\% & 100\% & 0\% & 0\% & 0\% \\
\rowcolor{Gray}
data exposure & 1\% & 0\% & 33\% & 33\% & 0\% & 33\% & 0\% & 33\% & 33\% & 33\% & 0\% \\
timing & 2\% & 25\% & 25\% & 0\% & 75\% & 0\% & 0\% & 75\% & 0\% & 25\% & 0\% \\
\rowcolor{Gray}
coding-bug & 2\% & 0\% & 0\% & 67\% & 33\% & 0\% & 0\% & 17\% & 0\% & 83\% & 0\% \\
front-running & 2\% & 0\% & 0\% & 80\% & 0\% & 20\% & 0\% & 100\% & 0\% & 0\% & 0\% \\
\rowcolor{Gray}
auditing and logging & 4\% & 0\% & 0\% & 0\% & 33\% & 44\% & 22\% & 33\% & 0\% & 56\% & 11\% \\
missing-logic & 1\% & 0\% & 0\% & 0\% & 67\% & 33\% & 0\% & 0\% & 0\% & 100\% & 0\% \\
\rowcolor{Gray}
cryptography & 0\% & 0\% & 0\% & 0\% & 100\% & 0\% & 0\% & 100\% & 0\% & 0\% & 0\% \\
documentation & 2\% & 0\% & 0\% & 0\% & 25\% & 50\% & 25\% & 0\% & 0\% & 75\% & 25\% \\
\rowcolor{Gray}
API inconsistency & 1\% & 0\% & 0\% & 0\% & 0\% & 100\% & 0\% & 0\% & 0\% & 100\% & 0\% \\
code-quality & 1\% & 0\% & 0\% & 0\% & 0\% & 100\% & 0\% & 0\% & 0\% & 100\% & 0\% \\
  \end{tabular}

  \vspace{0.1in}
  \caption{Severity and difficulty distributions for finding
    categories.  The second column shows what percent of all findings that
    category represents; the remaining columns are percentages
    \emph{within-category.}}
  \label{tab:overallcatpercent}
\end{table}

\begin{table}[t]
  \begin{tabular}{lrr|lrr}
Category & \% Dynamic & \% Static & Category & \% Dynamic & \% Static \\
    \hline
\rowcolor{Gray}    
data validation & 57\% & 22\% & logic & 0\% & 0\% \\
access controls & 50\% & 4\% & data exposure & 0\% & 0\% \\
\rowcolor{Gray}
    race condition & 6\% & 59\% & timing & 50\% & 25\% \\
numerics & 46\% & 69\% & coding-bug & 67\% & 50\% \\
\rowcolor{Gray}
    undefined behavior & 0\% & 31\% & front-running & 0\% & 0\% \\
patching & 17\% & 33\% & auditing and logging & 0\% & 38\% \\
\rowcolor{Gray}
    denial of service & 40\% & 0\% & missing-logic & 67\% & 0\% \\
authentication & 25\% & 0\% & cryptography & 0\% & 100\% \\
\rowcolor{Gray}
    reentrancy & 75\% & 100\% & documentation & 0\% & 0\% \\
error reporting & 29\% & 14\% & API inconsistency & 0\% & 0\% \\
\rowcolor{Gray}
    configuration & 0\% & 0\% & code-quality & 0\% & 67\% \\
  \end{tabular}

  \vspace{0.1in}
  \caption{Optimistic percentages of each category detectable by
    automated methods.}
  \label{tab:findway}
\end{table}

\subsubsection{Data Validation} Data validation covers the
large class of findings in which the core problem is that input received
from an untrusted source (e.g., arguments to a {\tt public} function
of a contract) is not properly vetted, with potentially harmful consequences (the type of harm varies widely).  Not only is this a frequently appearing problem, with more
than three times as many findings as the next most common category,
it is a \emph{serious} issue in many cases, with the largest absolute
number of high-low findings (10), and a fairly high percent of high-low
findings (11\%).  
Data validation can sometimes be detected
statically, by using taint to track unchecked user input to a
dangerous operation (e.g., an array
de-reference), but in many cases the consequences are not obviously problematic unless one understands a contract's purpose.
Ironically, the safer execution semantics of Solidity/EVM make
some problems that would clearly be security flaws in C or C++
harder to automatically detect.  In Solidity, it is not always incorrect to allow a user to provide an
array index: If the index is wrong, in many cases, the call will
simply revert, and there is no rule that contract code should
never revert.  From the point of view of a fuzzer or static analysis
tool, distinguishing bad reverts from intended ones is difficult
without guidance.  Automated static or dynamic
analysis to detect many of the instances of missing/incorrect data
validation identified in the audits would require
some user annotations, either in the form of properties or at least
annotating some functions or statements as not expected to revert, but
given that information, would likely prove effective.

\subsubsection{Access Controls}  Access control findings
describe cases where use of a legitimate operation of a contract should be
restricted to certain callers (the owner, minters, etc.), but
access control is either faulty or not implemented at all.  
Most often, access control findings are cases where access control is too permissive,
but nearly a third of these findings involve overly restrictive access control.  
While there are three times as many data validation
findings as access control findings, there are nearly as many high-low
findings for access control as for data validation.  One in four access control findings
is high-low, and 42\% of access control findings
are high severity.  In general, automatic detection of access control
problems without additional specification is often plausible.  In four
of our findings, it would suffice to check standard ERC20 token semantics,
enforce the {\tt paused} state for a contract, or assume that only
certain users should be able to cause self-destruction.
Cases where access controls are too restrictive would 
require additional specification, but, given that effort, are also often
likely to be handled well by property-based testing.

\subsubsection{Race Condition}  Race conditions are cases
in which the behavior of a contract depends (in an unintended
way) on an improperly restricted ordering of operations or events.  Often, the consequence of one particular
unexpected ordering is clearly incorrect.   The race condition
category had zero high-low findings, but was responsible for seven of
the 60 total high-severity findings across all audits.  The top three
categories (data validation, access controls, and race conditions)
made up over half of all high-severity findings.  A full 41\% of
race conditions are high severity.  Nearly half (nine) of the race
condition findings concern a known ERC20 issue \cite{erc20approve}, and could certainly be identified
automatically by a static analysis tool.  Due to the nature of many
blockchain race conditions, understanding the impact of the race would often
be hard for a dynamic analysis.

\subsubsection{Numerics} 
Numerics findings involve the
semantics of Solidity arithmetic: Most are overflow errors, some are
underflow errors, and a few involve precision losses.  These findings
also include cases where a ``safe math'' library or function is used,
so there is no actual overflow/underflow resulting in an incorrect
value, but the resulting revert causes problems. Three numerics
findings are high-low (23\%), and 31\% are high severity.  Rounding or
precision (six findings) and overflow (three findings) are the most
common numerics errors.  Many rounding and overflow problems
can likely be flagged
using static analysis, but to determine whether the
behavior is problematic would require custom properties.

\subsubsection{Undefined Behavior} 
The undefined behavior category includes cases where a contract relies on unspecified or
under-specified semantics of the Solidity language or the EVM, so
the actual semantic intent of the contract is either currently unclear or
may become so in the future. For instance, in Solidity, the evaluation order
of expressions in the same statement is not specified. Instead, it is only guaranteed 
that statements are executed in order. 
Three (23\%) of the undefined behavior
findings are high-low, and 31\% of undefined behavior findings are
high severity.  Undefined behavior is often easy to statically
detect.

\subsubsection{Patching} Patching findings concern flaws in
the process to upgrade or change contract behavior.  The immutability
of code on the blockchain requires the use of complex,
hard-to-get-right methods to allow changes.  Two (11\%) of the
patching findings are high-low, and 17\%  are high
severity. 
Many patching issues are complex environmental problems
that likely require human expertise, but some
common patterns of bad upgrade logic might be amenable to static
detection, and a dynamic analysis can detect that a contract is
broken after a faulty update.

\subsubsection{Denial of Service} 
Denial of service covers
findings that are not well described by another class (e.g., if lack of
data validation causes denial of service, we still classify it as data validation), and where the consequence of a flaw is either complete
shut-down of a contract or significant operational
inefficiency.  If we included all cases where denial of service is an important
potential consequence of a flaw, or even the only important
consequence, the category would be larger.  One denial
of service finding was high-low, and 20\% of findings were high
severity.  Most denial of service findings would require fairly complex custom properties specifying system
behavior, in part because ``simple'' denial of service due to some
less complex cause falls under another category.

\subsubsection{Authentication} Authentication findings specifically concern cases where the
mechanism used to determine identity or authorization is flawed, as
opposed to cases where the access rules are incorrect.  That is,
in authentication problems, the logic of who is allowed to
do what is correct, but the determination of ``who'' is flawed.  While
only one authentication finding is high-low, fully half of all
authentication problems are high severity; in fact, authentication is tied with
the infamous reentrancy problem in terms having the greatest percentage of high severity issues.  Three of the
observed authentication problems are highly idiosyncratic, and may not
even be automatically detectable with complex custom properties.  However, the remaining problem should be dynamically detectable
using ``off-the-shelf'' ERC20 token semantics properties.

\subsubsection{Reentrancy}  Reentrancy is a widely discussed
and investigated flaw in Ethereum smart contracts \cite{SurveyAttacks}.
In a reentrancy attack, a contract calls an external contract, before
``internal work'' (primarily state changes) is finished.  Through
some route, the external contract \emph{re-enters} code that expected
the internal work to be complete.  No reentrancy
problems detected in audits were high-low, but 50\% of the findings
were high severity.  Reentrancy is a serious problem, but, due to its
well-defined structure, is usually amenable to static and dynamic
detection.  In particular, static detection with relatively few false
positives is probably already possible using Slither, for most
important reentrancies.

\subsubsection{Error Reporting}  Error reporting findings
involve cases in which a contract does not properly report, propagate, or
handle error conditions.  There are no high-low error reporting
findings in the audits, but 29\% of error reporting findings are high
severity.  In some cases error reporting is a
difficulty category to capture without further specification, and specifying that errors should be reported or handled in a
certain way generally requires the same understanding that would have
produced correct code in the first place.  However, ERC20 semantics make some error reporting problems easy to automatically detect.  Incorrect error
\emph{propagation} is also usually statically detectable \cite{CindyLiblit};
however, this was not the type of error reporting problem discovered
in audits.

\subsubsection{Configuration} 
Configuration findings
generally describe cases in which a bad configuration may lead to bad behavior even when the contract itself is correct.
In smart contracts, this is often related to financial effects,
e.g., bad market/pricing parameters.
There are no high-low findings in this category, but 40\% of
findings are high priority.   Configuration problems are usually
fairly subtle, or even economic/financial in nature, and detection is likely to
rely on manual analysis.

\subsubsection{Logic}  
Logic findings describe incorrect
protocols or business logic, where the implementation is as
intended, but the reasoning behind the intention
is incorrect.  Somewhat surprisingly, this category has no
high-low findings, and only three fundamental logic flaws were
described in the audits.  One of the three logic flaws described was
high severity, however.  Based on the small number of findings it is
hard to guess how often custom properties might allow dynamic
detection of logic flaws. If the bad logic often leads to a violation
of the expected invariants of a contract, then it can be detected, but
if the fault is in the understanding of desirable invariants (which
may often be the case), manual inspection by another set of expert
eyes may be the only plausible detection method.

\subsubsection{Data Exposure}  Data exposure findings are
those in which information that should not be public is made public.
For instance, some smart contracts offer guarantees to users regarding  
the information about them stored on the blockchain. If an attacker can 
infer data about users by observing confirmed or unconfirmed transactions, 
then that is classified as a data exposure issue.
There are no high-low data exposure findings, but 33\%  are high severity. 
Most data exposure problems are
not likely to be amenable to automatic detection.

\subsubsection{Timing}  Timing findings concern cases (that
are neither race conditions nor front-running) where manipulation of
timing has negative consequences.  For the most part, these  findings involved
assuming intervals between events (especially blocks) that may not
hold in practice.  One of the four timing findings (the only high
severity one) was high-low.  Timing problems can be amenable to automated detection in that static or dynamic analysis
can certainly recognize when code depends on, for instance, the block
timestamp.

\subsubsection{Coding-Bug}  Coding-bug is a catch-all
category for problems that, whatever their consequences, amount to a
``typo'' in code, rather than a likely intentional error on a
developer's part.  Off-by-one loop bounds that do not traverse an
entire array are a simple example.  There were no high-low or
high-severity coding bugs in the smart contracts audited, which
suggests that the worst simple coding problems may be detected by existing
unit tests or human inspection of code, in the relatively small code bases of even larger smart contracts.  On the other hand, 67\% of
coding bugs were medium severity, the second-highest rate for that
severity; only one other class exceeded 41\% medium-severity findings.

\subsubsection{Front-Running} Front-running generalizes the
financial market concept of front-running, where a trader uses
advance non-public knowledge of a pending transaction to ``predict''
future prices and/or buy or sell before the pending state change. 
In smart contracts, this means that a contract 1) exposes information
about future state changes (especially to a ``market'') and
2) allows transactions that exploit this knowledge.  It is
both a timing and data exposure problem, but is assigned its own
category because the remedy is often different.  Front-running
is a well-known concern in smart contracts, but in fact no high-low or
even high-severity front-running problems were detected in our audits.  On
the other hand, front-running had the largest percent of
medium-severity findings (80\%), so it is not an insignificant
problem. 
Front-running, by its nature, is probably hard to detect
dynamically, and very hard to detect statically.

\subsubsection{Auditing and Logging}  Auditing and logging
findings describe inadequate or incorrect logging; in most cases incorrect or missing contract events.  There were no
high-low, high-severity, or medium-severity auditing or logging
findings.  If explicit checks for events are included in (automated)
testing, such problems can easily be detected, but if such checks are
included, the important events are also likely to be present and
correct, so this is not a great fit for dynamic analysis.
On the other hand, it is often easy to statically note when an
important state change is made but no event is associated
with it.

\subsubsection{Missing-Logic}  Missing-logic findings are
cases in which---rather than incorrect logic for handling a particular set
of inputs, or missing validation to exclude those inputs---there is a
correct way to handle inputs, but it is missing.  Structurally,
missing-logic means that code should add another branch to
handle a special case.  Interestingly, while this seems like
a potentially serious issue, there were no high-low or even medium-severity
missing-logic findings.  The ease of detecting missing logic with
custom properties depends  on the consequences of the omission; static analysis seems unlikely find most missing logic. 

\subsubsection{Cryptography}  Cryptography findings concern cases where incorrect or insufficient
cryptography is used.  In our smart contract audits, the one (low
severity, high difficulty) cryptography
finding concerned use of an improper pseudo-random number generator,
something a static analysis tool can often flag in the
blockchain context, where bad sources of randomness are fairly limited.

\subsubsection{Documentation}  Documentation findings
describe cases where the contract code is not incorrect, but there is
missing or erroneous documentation. As you would expect, this is
never a high- or even medium-severity issue, and is not amenable to
automated detection.

\subsubsection{API Inconsistency}  API inconsistencies are
cases in which a contract's individual functions are correct, but the
calling pattern or semantics of related functionalities differs in a
way likely to mislead a user and produce incorrect code calling
the contract.  All of these issues were informational, and while it is
conceivable that machine learning approaches could identify API
inconsistencies, it is not a low-hanging fruit for automated detection.

\subsubsection{Code-Quality}  Finally, code quality issues
have no semantic impact, but involve code that is hard to read or
maintain.  As expected, such issues are purely informational.
Code quality problems in general would seem to be highly amenable to
static
analysis, but not to dynamic analysis.

\subsection{Comparison to
  Non-Smart-Contract Audits}

It is interesting to compare the distribution of finding types for
smart contract audits to other security audits~\cite{reports}
performed by the same company.   Table \ref{otheraudits} compares smart contract audit
frequencies with those for a random sample of 15 non-smart contract audits,
with categories never present in smart contract audits or only present
in smart contract audits removed.



\begin{table}[t]
  \centering
  \begin{tabular}{l|r|r|r||l|r|r|r}
Category & \# & \% & Change & Category & \# & \% & Change\\
    \hline
\rowcolor{Gray}    
 data validation & 41 & 53\% & -17\% & patching & 6 & 8\% & -1\% \\
denial of service & 23 & 30\% & -26\% & authentication & 5 & 6\% & -4\% \\
\rowcolor{Gray}
configuration & 20 & 26\% & -24\% & timing & 4 & 5\% & -3\% \\
data exposure & 18 & 23\% & -22\% & numerics & 2 & 3\% & +3\% \\
\rowcolor{Gray}
access controls & 14 & 18\% & -8\% & auditing and logging & 2 & 3\% & +1\% \\
    cryptography & 12 & 16\% & -16\% & race condition & 1 & 1\% & +6\% \\
\rowcolor{Gray}
undefined behavior & 7 & 9\% & -4\% & error reporting & 1 & 1\% & +2\% \\
  \end{tabular}

  \vspace{0.1in}
  \caption{Most common finding categories in other audits.}
  \label{otheraudits}
\end{table}

The largest changes are categories of findings that are common in
other audits, but not common in smart contracts.  One of these, denial
of service, may be primarily due to the re-categorization of denial of
service findings with a clear relevance to another category in the
smart contract findings.  Changing the five findings whose type was
clarified back to denial of service still leaves a significant gap,
however.  This is likely due to the different nature of interactions
with the network in non-smart-contract code; in a sense, many denial
of service problems and solutions are delegated to the general
Ethereum blockchain, so individual contracts have less responsibility
and thus fewer problems.

A more general version of the same difference likely explains why configuration problems
are far less prevalent in smart contract code.  At heart, smart contracts are more specialized and focused,
and live in a simple environment (e.g., no OS/network interactions), so the
footprint of configurations, and thus possible mis-configurations, is
smaller.  Similarly, the temptation to roll your own cryptography in a
smart contract is much smaller.  For one thing, implementing any custom
cryptography in Solidity would be impractical enough to daunt even
those unwise enough to attempt it, and gas costs would be
prohibitive.  Data validation is also easier in a world where, for the
most part, transactions are the only inputs. Data exposure problems
are probably less common because it is  well understood that information on the
blockchain is public, so the amount of data that is presumed
unexposed is much smaller, or, in many cases, non-existent.


\subsection{Threats to Validity}

Contracts submitted for audit varied in their level of maturity; some assessments were performed on contracts essentially ready for release (or already released) that reflected the final stage of internal quality control processes.  Others were performed on much more preliminary implementations and designs.  This does not invalidate the findings, but some flaw types may be more prevalent in less polished contracts.  Of course, the primary threat to validity is that the data is all drawn from a set of 23 audits performed by one company over a period of about two years. We address this concern in Section \ref{sec:other}.

\section{Discussion: How to Find Flaws in Smart Contracts}

\subsection{Property-Based Testing and Symbolic Execution}

Property-based testing \cite{ClaessenH00,Hypothesis,TSTL} involves 1)
a user defining custom properties (usually, in practice, reachability
properties declaring certain system states or function return values
as ``bad''), and then 2) using either fuzzing or symbolic execution to attempt
to find inputs or call sequences violating the properties.  Some
variant of property-based testing is a popular approach to smart
contract analysis.
Automated testing with custom properties is both a significant
low-hanging fruit and anything but a panacea.  Of the 246 findings,
only 91 could be possibly labeled as detectable with 
user-defined properties, or with automated testing for standard
semantics of ERC20 tokens and other off-the-shelf dynamic checks.  On the other hand, 17 of the 27 most
important, high severity, low difficulty, findings, were plausibly
detectable using such properties.  While not effective for some
classes of problems, analysis using custom properties (and thus,
likely, dynamic rather than static analysis), might have detected over
60\% of the most important findings.  This mismatch in overall (37\%)
and high-low (63\%) percent of findings amenable to property-based
testing is likely due to the fact that categories almost never
detectable by automated testing---code quality, documentation, auditing
and logging---are seldom high-low, and those where it is most effective---data validation, access controls, and numerics---constitute a large
portion of the total set of high-low findings.  Also, intuition tells
us that if a finding has major detrimental consequences (high
severity) but is not extremely hard to exploit (low difficulty) this
is precisely the class of problems a set of key invariants plus
effective fuzzing or symbolic execution is suited to find.


\subsection{Static Analysis}

The full potential of static analysis is harder to estimate.  Four of
the issues in these findings were definitely detected using the
Slither static analysis tool, which has continued to add new detectors
and fix bugs since the majority of the audits were performed.  Of
these four issues, one was high severity, undetermined difficulty, a classic
reentrancy.  An additional four issues are certainly
detectable using Slither (these involve deletion of mappings, which
is also the root issue in one of the findings that was definitely detected by Slither).  Some of the overflow/underflow problems, as noted
above, might also be statically detectable if false positives are
allowed.   There are likely other individual findings amenable to
static analysis, but determining the practicality of such detection is
in some ways more difficult than with dynamic analysis using a
property-based specification.  The low-hanging fruit for static
analysis is general patterns of bad code, not reachability of a
complex bad state.  While some cases in which we speculate that a finding
is describable by a reachability property may not, in fact, prove
practical---current tools may have too much trouble generating a
transaction sequence demonstrating the problem---it is fairly easy
to determine that there is indeed an actual state of the contract that
can be identified with the finding.  Whether a finding falls into a
more general pattern not currently captured by, for instance, a Slither
detector, is harder to say, since the rate of false positives and
scalability of precision needed to identify a problem is very hard to
estimate.  Our conservative guess is that perhaps 65 of the 246
findings (26\%), and 9 of the
high-low findings (33\%), are plausibly detectable by static analysis.  While
these are lower percentages than for dynamic approaches, the
effort required is much, much lower: The dynamic analysis usually
depends on a user actually thinking of, and correctly implementing,
the right property, as well as a tool reaching the bad state.  For the
statically detectable problems, issues like those in
these findings would almost always be found just by running the static analysis tool.

\subsection{Unit Testing}

There was no additional unit testing as part of the
security audits performed.  It is therefore impossible to say how
effective adding unit tests would be in discovering flaws during
audits, based on this data.  However, it is possible to examine the
relationship between pre-existing unit tests and the audit results.
Fourteen of the contracts audited had what appeared to be
considerable unit tests; it is impossible to determine the quality of
these tests, but there was certainly quantity, and significant
development effort.  Two of the contracts had moderate unit tests; not
as good as the 14 contracts in the first category, but still
representing a serious effort to use unit testing.  Two contracts had
modest unit tests: non-trivial, but clearly far from complete tests.
Three had weak unit tests; technically there were unit tests, but 
they are practically of almost no value in checking the
correctness of the contract.  Finally, two contracts appeared to have
no unit tests at all.  Did the quantity of unit tests have an impact
on audit results?  If so, the impact was far from clear.  The contracts
that appeared to lack unit tests had nine and four findings,
respectively: fewer than most other contracts.  The largest mean
number of issues (11.5) was for contracts with modest unit tests, but
essentially the mean finding counts for considerable (11.1), moderate
(10.5), modest (11.5), and weak (11) unit tests were
indistinguishable.  Furthermore, restricting the analysis to counting only high-severity
findings also produces no significant correlation.  For total
findings, Kendall $\tau$ correlation is an extremely weak 0.09 ($p=0.61$
) indicating even this correlation is likely to be
pure chance.  For high-severity findings, the $\tau$ correlation drops
to 0.5 ($p=0.78$).  Note further that these
weak/unsupported correlations are \emph{in the ``wrong'' direction}.
 It seems fair to
say that even extensive unit tests are not the most effective way to
avoid the kind of problems found in high-quality
security audits.

\subsection{Manual Analysis}

With few exceptions, these findings demonstrate the effectiveness of
manual analysis.  Expert attention from experienced 
auditors can reveal serious problems even in well-tested code
bases.  While four of the audits produced no high-severity findings,
11 audits found three or more.  As far as we
can tell, all of the high-low severity issues were the result of manual
analysis alone, though there were recommendations for how to use tools
to detect/confirm correction in some cases.

\subsection{Recommendations}

The set of findings that could possibly be detected by \emph{either} dynamic
\emph{or} static analysis is slightly more than 50\%, and,
most importantly, includes 21 of the 27 high-low findings.  That is, making
generous assumptions about scalability, property-writing, and
willingness to wade through false positives, a skilled user of both
static and dynamic tools could detect more than three out of four
high-low issues.  Note that the use of both approaches is key: 61
findings overall and 12 high-low findings are likely to only be
detectable dynamically, while 35 findings, four of them high-low,
are likely to only by found using static analysis.

While static analysis alone is less powerful than manual audits or
dynamic analysis, the low effort, and thus high cost-benefit ratio, makes the use of all available
high-quality static analysis tools an obvious recommendation.  (Also, printers
and code understanding tools often provided by static analyzers
make manual audits more effective \cite{slitherpaper}.) Some of
the findings in these audits could have been easily detected by
developers using then-current versions of the best tools.

When 35\% of high-severity findings are not
likely to be detected even with considerable tool improvement and
manual effort to write correctness properties, it is implausible to
claim that tools will be a ``silver bullet'' for smart contract security.  It is difficult, at best, to imagine that nearly half of the
total findings and almost 25\% of the high-low findings would be detected even
with high-effort, high-expertise construction of custom properties and the
use of better-than-state-of-the-art dynamic and static analysis. Therefore, 
manual audits by external experts will remain a key part of serious
security and correctness efforts for smart contracts for the
foreseeable future.  

On the other hand, the gap between current
tool-based detection rates (very low) and our estimated upper limit on detection
rates (50\% of all issues, and over 75\% of the most
important issues) suggests that there is a large potential payoff
from improving state-of-the-art standards for analysis tools and putting more
effort into property-based testing.  The experience
of the security community using AFL, libFuzzer, and other tools also
suggests that there are ``missing'' findings.   The relatively immature
state of analysis tools when most of these audits were performed
likely means that \emph{bugs unlikely to be detected by human
reasoning were probably not detected}.  The effectiveness of fuzzing
in general suggests that such bugs likely exist in smart contracts as
well, especially since the most important target category of findings
for dynamic analyses, data validation, remains a major source of smart
contract findings.  In fact, a possible additional explanation for the
difference of 36\% data validation findings for smart contract audits
and 51\% for non-smart-contract audits could be that
non-smart-contract audits have access to more powerful fuzzers.
Eliminating the low-hanging fruit for automated tools will give
auditors more time to focus on the vulnerabilities that require
humans-in-the-loop and specialized skills.  Moreover, effort spent
writing custom properties is likely to pay off, even if dynamic
analysis tools are not yet good enough to produce a failing test. Just
understanding what invariants \emph{should} hold is often enough to alert a
human to a flaw.

Finally, while it is impossible to make strong claims based on a set
of only 23 audits, it seems likely that unit tests, even quite
substantial ones, do not provide an effective strategy for avoiding
the kinds of problems detected during audits.  Unit tests, of course,
have other important uses, and should be considered an essential part
of high-quality code development, but developer-constructed manual
unit tests may not really help detect high-severity security issues.
It does seem likely that the effort involved in writing high-quality
unit tests would be very helpful in dynamic analysis: Generalizing
from unit tests to invariants and properties for property-based
testing seems likely to be an effective way to detect some of what the
audits exposed.

\section{Audits From Other Companies}
\label{sec:other}

In order to partially validate our findings, we also performed an
analysis of  audits prepared by two other leading companies in the
field~\cite{michaelother}, ChainSecurity and ConsenSys Diligence.  While differences
in reporting standards and categorizations, and the fact that we do not
have access to unpublished reports (which could bias statistics), make
it difficult to analyze these results with the same confidence as our
own reports, the overall picture that emerged was broadly compatible
with our conclusions.  The assignment of findings to
semantically equivalent difficulties and severities, and the assessment of
potential for automated analysis methods, was performed by a completely
independent team.  The results summarized here are for 225 findings in
public reports for ChainSecurity and 168
from ConsenSys Diligence, over 19 and 18 audits, respectively.   Appendix B provides detailed results on these findings. 

First, the potential of automated methods is similar.  For ChainSecurity, 39\% of all issues were plausibly detectable by dynamic
analysis (e.g., property-based testing, possibly with a custom
property), and 22\% by automated static analysis.  For ConsenSys Diligence, those numbers were 41\% and 24\%.  Restricting our interest
to {\bf high-low} findings, the percentages were 67\% and 63\% for
dynamic analysis and 11\% and 38\% for static analysis, respectively.
Combining both methods, the potential detection rates were 51\% and
52\% for all findings, and 67\% and 75\% for high-low findings.  The
extreme similarity of these results to ours affirms that our results concerning detection methods are unlikely to be an
artifact of our audit methods or the specific set of contracts we audited.

Second, while the category frequencies were quite different than those in our
audits (e.g., more numerics and access controls, fewer data
validation findings), there were no new categories, and all of our
categories were present (though ChainSecurity found no race
conditions). Reentrancy was not, as previous literature might
lead one to suspect, a prominent source of high-low problems, or even 
a very common problem, and there
was only one high-low reentrancy. 

\section{Conclusions}

Understanding how best to protect high-value smart contracts against attackers
(and against serious errors by non-malicious users or the creators of
the contract) is difficult in the absence of information about the
actual problems found in high-value smart contracts by experienced
auditors using state-of-the-art technologies.  This paper presents a
wealth of empirical evidence to help smart-contract developers, security researchers, and security auditors improve their
understanding of the types of faults found in contracts, and the potential for
various methods to detect those faults.  Based on an in-depth examination of
23 paid smart contract audits performed by Trail of Bits, validated by
a more limited examination of public audits performed by
ChainSecurity and ConsenSys Diligence , we conclude that 1) the literature is
somewhat misleading with respect to the most important kinds of smart
contract flaws, which are more like flaws in other critical code than
one might think; 2) there is likely a large potential payoff in
making more effective use of automatic static and dynamic analyses to
detect the worst problems in smart contracts; 3) nonetheless, many key
issues will never be amenable to purely-automated or formal
approaches, and 4) high-quality unit tests alone do not provide effective protection
against serious contract flaws.  As future work, we plan to extend our
analysis of other companies' audits to include unit test quality, and examine issues that cut across findings categories,
such as the power of ERC20 standards to help find flaws.

\newpage

\bibliographystyle{plain}
\bibliography{contractbugs}

\begin{thebibliography}{10}

\bibitem{SurveyAttacks}
Nicola Atzei, Massimo Bartoletti, and Tiziana Cimoli.
\newblock A survey of attacks on {Ethereum} smart contracts {SoK}.
\newblock In {\em International Conference on Principles of Security and
  Trust}, pages 164--186, 2017.

\bibitem{measurepop}
Santiago Bragagnolo.
\newblock On contract popularity analysis.
\newblock
  \url{https://github.com/smartanvil/smartanvil.github.io/blob/master/\_posts/2018-03-14-on-contract-popularity-analysis.md}.

\bibitem{Brent2018VandalAS}
Lexi Brent and et~al.
\newblock Vandal: A scalable security analysis framework for smart contracts.
\newblock {\em CoRR}, abs/1809.03981, 2018.

\bibitem{buterin2013whitepaper}
Vitalik Buterin.
\newblock Ethereum: A next-generation smart contract and decentralized
  application platform.
\newblock \url{https://github.com/ethereum/wiki/wiki/White-Paper}, 2013.

\bibitem{ClaessenH00}
Koen Claessen and John Hughes.
\newblock {QuickCheck}: a lightweight tool for random testing of {Haskell}
  programs.
\newblock In {\em International Conference on Functional Programming {(ICFP)}},
  pages 268--279, 2000.

\bibitem{mythril-code}
ConsenSys.
\newblock Mythril: a security analysis tool for ethereum smart contracts.
\newblock \url{https://github.com/ConsenSys/mythril-classic}, 2017.

\bibitem{dika2017ethereum}
Ardit Dika.
\newblock Ethereum smart contracts: Security vulnerabilities and security
  tools.
\newblock Master's thesis, NTNU, 2017.

\bibitem{smartanvil}
St{\'e}phane Ducasse, Henrique Rocha, Santiago Bragagnolo, Marcus Denker, and
  Cl{\'e}ment Francomme.
\newblock Smartanvil: Open-source tool suite for smart contract analysis.
\newblock Technical Report hal-01940287, {HAL}, 2019.

\bibitem{slitherpaper}
Josselin Feist, Gustavo Greico, and Alex Groce.
\newblock Slither: A static analysis framework for smart contracts.
\newblock In {\em International Workshop on Emerging Trends in Software
  Engineering for Blockchain}, 2019.

\bibitem{ethertrust}
Ilya Grishchenko, Matteo Maffei, and Clara Schneidewind.
\newblock Ethertrust: Sound static analysis of ethereum bytecode, 2018.

\bibitem{grishchenko2018semantic}
Ilya Grishchenko, Matteo Maffei, and Clara Schneidewind.
\newblock A semantic framework for the security analysis of ethereum smart
  contracts.
\newblock arXiv:1802.08660, 2018.
\newblock Accessed:2018-03-12.

\bibitem{TSTL}
Josie Holmes, Alex Groce, Jervis Pinto, Pranjal Mittal, Pooria Azimi, Kevin
  Kellar, and James O'Brien.
\newblock {TSTL:} the template scripting testing language.
\newblock {\em International Journal on Software Tools for Technology
  Transfer}, 20(1):57--78, 2018.

\bibitem{teether}
Johannes Krupp and Christian Rossow.
\newblock teether: Gnawing at ethereum to automatically exploit smart
  contracts.
\newblock In {\em USENIX Security )}, 2018.

\bibitem{oyente}
Loi Luu, Duc-Hiep Chu, Hrishi Olickel, Prateek Saxena, and Aquinas Hobor.
\newblock Making smart contracts smarter.
\newblock CCS '16, 2016.

\bibitem{Hypothesis}
David~R. MacIver.
\newblock Hypothesis: Test faster, fix more.
\newblock \url{http://hypothesis.works/}, March 2013.

\bibitem{Mense}
Alexander Mense and Markus Flatscher.
\newblock Security vulnerabilities in ethereum smart contracts.
\newblock In {\em Proceedings of the 20th International Conference on
  Information Integration and Web-based Applications \& Services}, iiWAS2018,
  pages 375--380, New York, NY, USA, 2018. ACM.

\bibitem{manticorepaper}
Mark Mossberg, Felipe Manzano, Eric Hennenfent, Alex Groce, Gustavo Greico,
  Josselin Feist, Trent Brunson, , and Artem Dinaburg.
\newblock Manticore: A user-friendly symbolic execution framework for binaries
  and smart contracts.
\newblock In {\em IEEE/ACM International Conference on Automated Software
  Engineering}.
\newblock accepted for publication.

\bibitem{maian}
Ivica Nikolic, Aashish Kolluri, Ilya Sergey, Prateek Saxena, and Aquinas Hobor.
\newblock Finding the greedy, prodigal, and suicidal contracts at scale.
\newblock In {\em ACSAC}, 2018.

\bibitem{DoesAnyoneCare}
Daniel Perez and Benjamin Livshits.
\newblock Smart contract vulnerabilities: Does anyone care?
\newblock 2019.

\bibitem{DAO}
{Phil Daian }.
\newblock Analysis of the dao exploit.
\newblock \url{http://hackingdistributed.com/2016/06/18/analysis-of-the-
  dao-exploit/}, June 18, 2016 (acceded on Jan 10, 2019).

\bibitem{CindyLiblit}
Cindy Rubio{-}Gonz{\'{a}}lez, Haryadi~S. Gunawi, Ben Liblit, Remzi~H.
  Arpaci{-}Dusseau, and Andrea~C. Arpaci{-}Dusseau.
\newblock Error propagation analysis for file systems.
\newblock In {\em {ACM} {SIGPLAN} Conference on Programming Language Design and
  Implementation ({PLDI})}, pages 270--280, 2009.

\bibitem{smartcheck}
Tikhomirov S. and et~al.
\newblock Smartcheck: Static analysis of ethereum smart contracts.
\newblock WETSEB, 2018.

\bibitem{ExploreAttackSurface}
Muhammad Saad, Jeffrey Spaulding, Laurent Njilla, Charles Kamhoua, Sachin
  Shetty, DaeHun Nyang, and Aziz Mohaisen.
\newblock Exploring the attack surface of blockchain: A systematic overview.
\newblock {\em arXiv preprint arXiv:1904.03487}, 2019.

\bibitem{spank}
{SpankChain}.
\newblock We got spanked: What we know so far.
\newblock \url{https://medium.com/spankchain/we-got-spanked-what-we-know
  -so-far-d5ed3a0f38fe}, Oct 8, 2018 (acceded on Jan 10, 2019).

\bibitem{manticore-code}
{Trail of Bits}.
\newblock Manticore: Symbolic execution for humans.
\newblock \url{https://github.com/trailofbits/manticore}, 2017.

\bibitem{echidna-code}
{Trail of Bits}.
\newblock Echidna: Ethereum fuzz testing framework.
\newblock \url{https://github.com/trailofbits/echidna}, 2018.

\bibitem{michaelother}
{Trail of Bits}.
\newblock Analysis of external audits.
\newblock \url{https://github.com/trailofbits/publications/tree/master/datasets
  /smart\_contract\_audit\_findings/other\_audit\_sources}, 2019.

\bibitem{contractrepo}
{Trail of Bits}.
\newblock Smart contract audit findings.
\newblock \url{https://github.com/trailofbits/publications/tree/master/datasets
  /smart\_contract\_audit\_findings}, 2019.

\bibitem{reports}
{Trail of Bits}.
\newblock Trail of bits security reviews.
\newblock \url{https://github.com/trailofbits/publications\#security-reviews},
  2019.

\bibitem{securify}
Petar Tsankov, Andrei Dan, Dana Drachsler-Cohen, Arthur Gervais, Florian
  B\"{u}nzli, and Martin Vechev.
\newblock Securify: Practical security analysis of smart contracts.
\newblock CCS '18, 2018.

\bibitem{erc20approve}
Github user: 3sGgpQ8H.
\newblock Attack vector on {ERC20} {API}.
\newblock
  \url{https://github.com/ethereum/EIPs/issues/20\#issuecomment-263524729}.

\bibitem{wood2014yellow}
Gavin Wood.
\newblock Ethereum: a secure decentralised generalised transaction ledger.
\newblock \url{http://gavwood.com/paper.pdf}, 2014.

\end{thebibliography}

\newpage

\section*{Appendix A: Raw Counts for Finding Categories}
\label{sec:exact}

This table provides exact counts for categories, and severities within categories, for our analysis.

\noindent  \begin{tabular}{lr|r|rrrrr|rrrr}
& & & \multicolumn{5}{c}{Severity} & \multicolumn{4}{|c}{Difficulty} \\
Category & \# & High-Low & High & Med. & Low & Info. & Und. & High & Med. & Low & Und.\\
             \hline
\rowcolor{Gray}             
data validation & 89 & 10 & 19 & 32 & 21 & 12 & 5 & 24 & 14 & 49 & 2 \\
access controls & 24 & 6 & 10 & 6 & 3 & 5 & 0 & 8 & 3 & 13 & 0 \\
\rowcolor{Gray}
             race condition & 17 & 0 & 7 & 7 & 1 & 2 & 0 & 17 & 0 & 0 & 0 \\
numerics & 13 & 3 & 4 & 3 & 5 & 1 & 0 & 4 & 1 & 8 & 0 \\
\rowcolor{Gray}
             undefined behavior & 13 & 3 & 4 & 2 & 4 & 1 & 2 & 2 & 1 & 10 & 0 \\
patching & 18 & 2 & 3 & 2 & 7 & 5 & 1 & 1 & 2 & 11 & 4 \\
\rowcolor{Gray}
             denial of service & 10 & 1 & 2 & 3 & 3 & 2 & 0 & 5 & 0 & 4 & 1 \\
authentication & 4 & 1 & 2 & 1 & 1 & 0 & 0 & 2 & 0 & 2 & 0 \\
\rowcolor{Gray}
             reentrancy & 4 & 0 & 2 & 1 & 1 & 0 & 0 & 2 & 1 & 0 & 1 \\
error reporting & 7 & 0 & 2 & 1 & 0 & 4 & 0 & 3 & 2 & 2 & 0 \\
\rowcolor{Gray}
             configuration & 5 & 0 & 2 & 0 & 1 & 1 & 1 & 3 & 1 & 1 & 0 \\
logic & 3 & 0 & 1 & 1 & 1 & 0 & 0 & 3 & 0 & 0 & 0 \\
\rowcolor{Gray}
             data exposure & 3 & 0 & 1 & 1 & 0 & 1 & 0 & 1 & 1 & 1 & 0 \\
timing & 4 & 1 & 1 & 0 & 3 & 0 & 0 & 3 & 0 & 1 & 0 \\
\rowcolor{Gray}
             coding-bug & 6 & 0 & 0 & 4 & 2 & 0 & 0 & 1 & 0 & 5 & 0 \\
front-running & 5 & 0 & 0 & 4 & 0 & 1 & 0 & 5 & 0 & 0 & 0 \\
\rowcolor{Gray}
             auditing and logging & 9 & 0 & 0 & 0 & 3 & 4 & 2 & 3 & 0 & 5 & 1 \\
missing-logic & 3 & 0 & 0 & 0 & 2 & 1 & 0 & 0 & 0 & 3 & 0 \\
\rowcolor{Gray}
             cryptography & 1 & 0 & 0 & 0 & 1 & 0 & 0 & 1 & 0 & 0 & 0 \\
documentation & 4 & 0 & 0 & 0 & 1 & 2 & 1 & 0 & 0 & 3 & 1 \\
\rowcolor{Gray}
             API inconsistency & 2 & 0 & 0 & 0 & 0 & 2 & 0 & 0 & 0 & 2 & 0 \\
code-quality & 2 & 0 & 0 & 0 & 0 & 2 & 0 & 0 & 0 & 2 & 0 \\
\hline
Total & 246 & 27 & 60 & 68 & 60 & 46 & 12 & 88 & 26 & 122 & 10 \\
  \end{tabular}

\newpage
 
\section*{Appendix B: Detailed Results for ChainSecurity and ConsenSys
  Diligence Smart
  Contract Audits}
\label{sec:othercompanydetails}

The process for analyzing findings in other companies' audits involved
1) mapping the category of the finding to our set, which was not
always simple or obvious, and 2) translating a different formulation
of worst-case impact and probability estimation into our high-low
severity and high-low difficulty schemes. For information on the original 
categorizations of issues (using a different severity and likelihood 
scheme), see the full data set online \cite{michaelother}.  A potential source of bias in these results is that we do 
not know the results for non-public audits for these companies; for 
our own audits, there was no obvious difference between public and 
non-public audits, however.  Due to the lack of access to
source code versions associated with audits, we were unfortunately unable to
correlate unit test quality at time of audit with issue counts for
external audits.

Note that both companies reported a large number of code
quality issues that would not have been considered findings at all in our own
audits, but simply noted in a Code Quality appendix to an audit
report.  We removed 66 and 168 such relatively trivial
(``lint-like'') findings,
respectively, for ChainSecurity and ConsenSys Diligence; including these would greatly
increase the counts for informational issues and the code-quality
category.

The first two tables show severity and difficulty distributions for 
finding categories for other company audits, as in Table
\ref{tab:overallcatpercent}.  In all cases, the first table in each
pair of tables is for ChainSecurity, and the second is for ConsenSys Diligence.

\noindent  \begin{tabular}{lr|r|rrrrr|rrrr}
& & & \multicolumn{5}{c}{Severity} & \multicolumn{4}{|c}{Difficulty} \\
Category & \% & High-Low & High & Med. & Low & Info. & Und. & High & Med. & Low & Und.\\
             \hline
\rowcolor{Gray}             
access controls & 24\% & 8\% & 28\% & 21\% & 45\% & 6\% & 0\% & 40\% & 26\% & 34\% & 0\% \\
data validation & 14\% & 3\% & 19\% & 28\% & 47\% & 6\% & 0\% & 47\% & 9\% & 44\% & 0\% \\
\rowcolor{Gray}
             logic & 6\% & 7\% & 36\% & 50\% & 14\% & 0\% & 0\% & 29\% & 50\% & 21\% & 0\% \\
numerics & 9\% & 0\% & 10\% & 15\% & 75\% & 0\% & 0\% & 40\% & 20\% & 40\% & 0\% \\
\rowcolor{Gray}
             denial of service & 5\% & 0\% & 17\% & 25\% & 58\% & 0\% & 0\% & 67\% & 33\% & 0\% & 0\% \\
configuration & 3\% & 14\% & 29\% & 29\% & 43\% & 0\% & 0\% & 71\% & 0\% & 29\% & 0\% \\
\rowcolor{Gray}
             authentication & 2\% & 0\% & 50\% & 25\% & 25\% & 0\% & 0\% & 0\% & 75\% & 25\% & 0\% \\
coding-bug & 2\% & 20\% & 40\% & 0\% & 60\% & 0\% & 0\% & 20\% & 0\% & 80\% & 0\% \\
\rowcolor{Gray}
             missing-logic & 4\% & 0\% & 13\% & 13\% & 63\% & 13\% & 0\% & 25\% & 0\% & 75\% & 0\% \\
cryptography & 1\% & 50\% & 50\% & 50\% & 0\% & 0\% & 0\% & 50\% & 0\% & 50\% & 0\% \\
\rowcolor{Gray}
             patching & 7\% & 0\% & 7\% & 0\% & 73\% & 20\% & 0\% & 87\% & 13\% & 0\% & 0\% \\
reentrancy & 2\% & 0\% & 20\% & 0\% & 80\% & 0\% & 0\% & 80\% & 0\% & 20\% & 0\% \\
\rowcolor{Gray}
             documentation & 4\% & 0\% & 13\% & 0\% & 50\% & 38\% & 0\% & 13\% & 13\% & 63\% & 0\% \\
data exposure & 0\% & 0\% & 100\% & 0\% & 0\% & 0\% & 0\% & 100\% & 0\% & 0\% & 0\% \\
\rowcolor{Gray}
             timing & 5\% & 0\% & 0\% & 27\% & 64\% & 9\% & 0\% & 45\% & 27\% & 27\% & 0\% \\
front-running & 2\% & 0\% & 0\% & 25\% & 75\% & 0\% & 0\% & 75\% & 25\% & 0\% & 0\% \\
\rowcolor{Gray}
             auditing and logging & 3\% & 0\% & 0\% & 14\% & 29\% & 57\% & 0\% & 14\% & 0\% & 86\% & 0\% \\
error reporting & 2\% & 0\% & 0\% & 25\% & 50\% & 25\% & 0\% & 0\% & 25\% & 75\% & 0\% \\
\rowcolor{Gray}
             undefined behavior & 1\% & 0\% & 0\% & 50\% & 50\% & 0\% & 0\% & 50\% & 50\% & 0\% & 0\% \\
API-inconsistency & 2\% & 0\% & 0\% & 0\% & 100\% & 0\% & 0\% & 40\% & 0\% & 60\% & 0\% \\
\rowcolor{Gray}
             code-quality & 3\% & 0\% & 0\% & 0\% & 50\% & 50\% & 0\% & 0\% & 0\% &
                                                                       83\%
                                                     & 0\% \\
race condition & 0\% & N/A & N/A & N/A & N/A & N/A & N/A & N/A & N/A &
                                                                       N/A
                                                     & N/A\\
  \end{tabular}
  
\noindent  \begin{tabular}{lr|r|rrrrr|rrrr}
& & & \multicolumn{5}{c}{Severity} & \multicolumn{4}{|c}{Difficulty} \\
Category & \% & High-Low & High & Med. & Low & Info. & Und. & High & Med. & Low & Und.\\
             \hline
\rowcolor{Gray}             
access controls & 10\% & 0\% & 35\% & 12\% & 47\% & 6\% & 0\% & 29\% & 29\% & 41\% & 0\% \\
configuration & 10\% & 6\% & 25\% & 13\% & 56\% & 0\% & 0\% & 56\% & 6\% & 31\% & 6\% \\
\rowcolor{Gray}
             front-running & 4\% & 14\% & 57\% & 14\% & 29\% & 0\% & 0\% & 71\% & 14\% & 14\% & 0\% \\
reentrancy & 4\% & 14\% & 43\% & 43\% & 14\% & 0\% & 0\% & 57\% & 14\% & 29\% & 0\% \\
\rowcolor{Gray}
             coding-bug & 6\% & 10\% & 30\% & 10\% & 50\% & 10\% & 0\% & 20\% & 10\% & 70\% & 0\% \\
logic & 8\% & 8\% & 15\% & 31\% & 54\% & 0\% & 0\% & 15\% & 23\% & 62\% & 0\% \\
\rowcolor{Gray}
             numerics & 13\% & 5\% & 10\% & 14\% & 71\% & 5\% & 0\% & 52\% & 24\% & 24\% & 0\% \\
data validation & 6\% & 0\% & 10\% & 20\% & 70\% & 0\% & 0\% & 50\% & 20\% & 30\% & 0\% \\
\rowcolor{Gray}
             API inconsistency & 2\% & 0\% & 25\% & 25\% & 50\% & 0\% & 0\% & 25\% & 0\% & 75\% & 0\% \\
cryptography & 1\% & 50\% & 50\% & 50\% & 0\% & 0\% & 0\% & 50\% & 0\% & 50\% & 0\% \\
\rowcolor{Gray}
             error reporting & 3\% & 20\% & 20\% & 0\% & 80\% & 0\% & 0\% & 20\% & 0\% & 80\% & 0\% \\
timing & 2\% & 0\% & 25\% & 0\% & 75\% & 0\% & 0\% & 50\% & 0\% & 50\% & 0\% \\
\rowcolor{Gray}
             race condition & 1\% & 0\% & 100\% & 0\% & 0\% & 0\% & 0\% & 100\% & 0\% & 0\% & 0\% \\
missing-logic & 11\% & 0\% & 0\% & 26\% & 68\% & 5\% & 0\% & 0\% & 11\% & 84\% & 0\% \\
\rowcolor{Gray}
             authentication & 1\% & 0\% & 0\% & 100\% & 0\% & 0\% & 0\% & 50\% & 50\% & 0\% & 0\% \\
denial of service & 2\% & 0\% & 0\% & 67\% & 0\% & 0\% & 0\% & 0\% & 33\% & 33\% & 0\% \\
\rowcolor{Gray}
             documentation & 2\% & 0\% & 0\% & 33\% & 33\% & 33\% & 0\% & 67\% & 0\% & 33\% & 0\% \\
data exposure & 1\% & 0\% & 0\% & 100\% & 0\% & 0\% & 0\% & 0\% & 0\% & 100\% & 0\% \\
\rowcolor{Gray}
             code-quality & 7\% & 0\% & 0\% & 0\% & 82\% & 9\% & 0\% & 45\% & 9\% & 36\% & 0\% \\
patching & 3\% & 0\% & 0\% & 0\% & 80\% & 20\% & 0\% & 80\% & 0\% & 20\% & 0\% \\
\rowcolor{Gray}
             undefined behavior & 1\% & 0\% & 0\% & 0\% & 50\% & 50\% & 0\% & 0\% & 0\% & 0\% & 100\% \\
auditing and logging & 3\% & 0\% & 0\% & 0\% & 0\% & 100\% & 0\% & 0\% & 0\% & 100\% & 0\% \\
  \end{tabular}

\newpage
The next two tables show absolute severity and difficulty counts for 
finding categories for other company audits, as in Appendix A.

\noindent  \begin{tabular}{lr|r|rrrrr|rrrr}
& & & \multicolumn{5}{c}{Severity} & \multicolumn{4}{|c}{Difficulty} \\
Category & \# & High-Low & High & Med. & Low & Info. & Und. & High & Med. & Low & Und.\\
             \hline
\rowcolor{Gray}             
access controls & 53 & 4 & 15 & 11 & 24 & 3 & 0 & 21 & 14 & 18 & 0 \\
data validation & 32 & 1 & 6 & 9 & 15 & 2 & 0 & 15 & 3 & 14 & 0 \\
\rowcolor{Gray}
             logic & 14 & 1 & 5 & 7 & 2 & 0 & 0 & 4 & 7 & 3 & 0 \\
numerics & 20 & 0 & 2 & 3 & 15 & 0 & 0 & 8 & 4 & 8 & 0 \\
\rowcolor{Gray}
             denial of service & 12 & 0 & 2 & 3 & 7 & 0 & 0 & 8 & 4 & 0 & 0 \\
configuration & 7 & 1 & 2 & 2 & 3 & 0 & 0 & 5 & 0 & 2 & 0 \\
\rowcolor{Gray}
             authentication & 4 & 0 & 2 & 1 & 1 & 0 & 0 & 0 & 3 & 1 & 0 \\
coding-bug & 5 & 1 & 2 & 0 & 3 & 0 & 0 & 1 & 0 & 4 & 0 \\
\rowcolor{Gray}
             missing-logic & 8 & 0 & 1 & 1 & 5 & 1 & 0 & 2 & 0 & 6 & 0 \\
cryptography & 2 & 1 & 1 & 1 & 0 & 0 & 0 & 1 & 0 & 1 & 0 \\
\rowcolor{Gray}
             patching & 15 & 0 & 1 & 0 & 11 & 3 & 0 & 13 & 2 & 0 & 0 \\
reentrancy & 5 & 0 & 1 & 0 & 4 & 0 & 0 & 4 & 0 & 1 & 0 \\
\rowcolor{Gray}
             documentation & 8 & 0 & 1 & 0 & 4 & 3 & 0 & 1 & 1 & 5 & 0 \\
data exposure & 1 & 0 & 1 & 0 & 0 & 0 & 0 & 1 & 0 & 0 & 0 \\
\rowcolor{Gray}
             timing & 11 & 0 & 0 & 3 & 7 & 1 & 0 & 5 & 3 & 3 & 0 \\
front-running & 4 & 0 & 0 & 1 & 3 & 0 & 0 & 3 & 1 & 0 & 0 \\
\rowcolor{Gray}
             auditing and logging & 7 & 0 & 0 & 1 & 2 & 4 & 0 & 1 & 0 & 6 & 0 \\
error reporting & 4 & 0 & 0 & 1 & 2 & 1 & 0 & 0 & 1 & 3 & 0 \\
\rowcolor{Gray}
             undefined behavior & 2 & 0 & 0 & 1 & 1 & 0 & 0 & 1 & 1 & 0 & 0 \\
API inconsistency & 5 & 0 & 0 & 0 & 5 & 0 & 0 & 2 & 0 & 3 & 0 \\
\rowcolor{Gray}
             code-quality & 6 & 0 & 0 & 0 & 3 & 3 & 0 & 0 & 0 & 5 & 0 \\
\hline
Total & 225 & 9 & 42 & 45 & 117 & 21 & 0 & 96 & 44 & 83 & 0 \\

           \end{tabular}

\noindent  \begin{tabular}{lr|r|rrrrr|rrrr}
& & & \multicolumn{5}{c}{Severity} & \multicolumn{4}{|c}{Difficulty} \\
Category & \# & High-Low & High & Med. & Low & Info. & Und. & High & Med. & Low & Und.\\
             \hline
\rowcolor{Gray}             
access controls & 17 & 0 & 6 & 2 & 8 & 1 & 0 & 5 & 5 & 7 & 0 \\
configuration & 16 & 1 & 4 & 2 & 9 & 0 & 0 & 9 & 1 & 5 & 1 \\
\rowcolor{Gray}
             front-running & 7 & 1 & 4 & 1 & 2 & 0 & 0 & 5 & 1 & 1 & 0 \\
reentrancy & 7 & 1 & 3 & 3 & 1 & 0 & 0 & 4 & 1 & 2 & 0 \\
\rowcolor{Gray}
             coding-bug & 10 & 1 & 3 & 1 & 5 & 1 & 0 & 2 & 1 & 7 & 0 \\
logic & 13 & 1 & 2 & 4 & 7 & 0 & 0 & 2 & 3 & 8 & 0 \\
\rowcolor{Gray}
             numerics & 21 & 1 & 2 & 3 & 15 & 1 & 0 & 11 & 5 & 5 & 0 \\
data validation & 10 & 0 & 1 & 2 & 7 & 0 & 0 & 5 & 2 & 3 & 0 \\
\rowcolor{Gray}
             API inconsistency & 4 & 0 & 1 & 1 & 2 & 0 & 0 & 1 & 0 & 3 & 0 \\
cryptography & 2 & 1 & 1 & 1 & 0 & 0 & 0 & 1 & 0 & 1 & 0 \\
\rowcolor{Gray}
             error reporting & 5 & 1 & 1 & 0 & 4 & 0 & 0 & 1 & 0 & 4 & 0 \\
timing & 4 & 0 & 1 & 0 & 3 & 0 & 0 & 2 & 0 & 2 & 0 \\
\rowcolor{Gray}
             race condition & 1 & 0 & 1 & 0 & 0 & 0 & 0 & 1 & 0 & 0 & 0 \\
missing-logic & 19 & 0 & 0 & 5 & 13 & 1 & 0 & 0 & 2 & 16 & 0 \\
\rowcolor{Gray}
             authentication & 2 & 0 & 0 & 2 & 0 & 0 & 0 & 1 & 1 & 0 & 0 \\
denial of service & 3 & 0 & 0 & 2 & 0 & 0 & 0 & 0 & 1 & 1 & 0 \\
\rowcolor{Gray}
             documentation & 3 & 0 & 0 & 1 & 1 & 1 & 0 & 2 & 0 & 1 & 0 \\
data exposure & 1 & 0 & 0 & 1 & 0 & 0 & 0 & 0 & 0 & 1 & 0 \\
\rowcolor{Gray}
             code-quality & 11 & 0 & 0 & 0 & 9 & 1 & 0 & 5 & 1 & 4 & 0 \\
patching & 5 & 0 & 0 & 0 & 4 & 1 & 0 & 4 & 0 & 1 & 0 \\
\rowcolor{Gray}
             undefined behavior & 2 & 0 & 0 & 0 & 1 & 1 & 0 & 0 & 0 & 0 & 2 \\
auditing and logging & 5 & 0 & 0 & 0 & 0 & 5 & 0 & 0 & 0 & 5 & 0 \\
\hline
Total & 168 & 8 & 30 & 31 & 91 & 13 & 0 & 61 & 24 & 77 & 3 \\

\end{tabular}

The final two tables report the estimated automated dynamic and static
analysis detection potential for the categories in the other
companies' audits.
  
\noindent \begin{tabular}{lrr|lrr}
Category & \% Dynamic & \% Static & Category & \% Dynamic & \% Static
            \\
            \hline
\rowcolor{Gray}            
access controls & 43\% & 6\% & reentrancy & 60\% & 100\%\\
data validation & 31\% & 13\% & documentation & 0\% & 0\%\\
\rowcolor{Gray}
            logic & 50\% & 7\% & data exposure & 0\% & 0\% \\
numerics & 80\% & 55\% & timing & 36\% & 36\%\\
\rowcolor{Gray}
            denial of service & 33\% & 25\% & front-running & 0\% & 0\%\\
configuration & 29\% & 0\% & auditing and logging & 0\% & 0\%\\
\rowcolor{Gray}
            authentication & 50\% & 25\% & error reporting & 100\% & 25\%\\
coding-bug & 100\% & 40\% & undefined behavior & 0\% & 50\%\\
\rowcolor{Gray}
            missing-logic & 63\% & 0\% & API-inconsistency & 20\% & 20\%\\
cryptography & 50\% & 0\% & code-quality & 0\% & 17\%\\
\rowcolor{Gray}
            patching & 7\% & 73\% & race condition & N/A & N/A\\
          \end{tabular}

 \noindent \begin{tabular}{lrr|lrr}
Category & \% Dynamic & \% Static & Category & \% Dynamic & \% Static
            \\
             \hline
\rowcolor{Gray}             
access controls & 18\% & 12\% &timing & 25\% & 0\%\\
configuration & 31\% & 25\% &race condition & 0\% & 0\%\\
\rowcolor{Gray}
             front-running & 0\% & 0\%  &missing-logic & 47\% & 0\%\\
reentrancy & 100\% & 71\%  &authentication & 50\% & 0\%\\
\rowcolor{Gray}
             coding-bug & 50\% & 10\% &denial of service & 67\% & 0\%\\
logic & 62\% & 8\% &documentation & 0\% & 0\%\\
\rowcolor{Gray}
             numerics & 95\% & 71\% &data exposure & 0\% & 0\%\\
data validation & 20\% & 10\% &code-quality & 9\% & 45\%\\
\rowcolor{Gray}
             API inconsistency & 50\% & 25\% &patching & 0\% & 60\%\\
cryptography & 100\% & 0\% &undefined behavior & 0\% & 50\%\\
\rowcolor{Gray}
             error reporting & 20\% & 40\% &auditing and logging & 0\% & 0\%\\

           \end{tabular}

\end{document}